\documentclass[a4paper,12pt]{article}
\usepackage{cite}
\newcommand{\be}{\begin{equation}}
\newcommand{\ee}{\end{equation}}
\newcommand{\bea}{\begin{eqnarray}}

\newcommand{\eea}{\end{eqnarray}}

\begin{document}
%%%%%%%%%%%%%%%%%%%%

\title{
%%%%%%%%%%%%%%%%%
%\begin{flushright}
%{\small UB-ECM-PF-04/33}\\
%\end{flushright}\vspace{1.5cm}
%%%%%%%%%%%%%%%%
On Hawking Radiation as Tunneling with Back-Reaction}

\author{A.J.M. Medved$\;^a$ and
Elias C. Vagenas$\;^b$ \\ \\
$\;^a$~School of Mathematics, Statistics and Computer Science \\
Victoria University of Wellington \\
PO Box 600, Wellington, New Zealand \\
E-Mail: joey.medved@mcs.vuw.ac.nz \\ \\
$\;^b$~Nuclear and Particle Physics Section\\
Physics Department\\
University of Athens\\
GR-15771, Athens, Greece\\
E-Mail: evagenas@phys.uoa.gr\\ \\ }

\maketitle
\begin{abstract}

Recently, Angheben {\it et. al.} \cite{Van} have presented a refined method
for calculating the (tree-level) black hole temperature by way of the 
tunneling paradigm.
In the current letter, we  demonstrate how  their formalism
can be suitably adapted to accommodate  the (higher-order) effects of
the gravitational back-reaction.

\end{abstract}

\newpage

\section*{}

\par
The tunneling paradigm \cite{KW,PW,EK,SP}~\footnote{And see 
\cite{Van} for an exhaustive list of references.} provides
an intuitively simple but physically rich framework for quantitatively
describing the process of black hole (Hawking) radiation \cite{Haw}. 
The basic idea is that the radiative process begins, just inside
the black hole horizon, with
the  quantum inducement  of a pair-production event. 
The spontaneously created particles
are then capable --- thanks to further quantum effects ---  
of traveling  along classically 
forbidden trajectories. In particular, the positive-energy particle
can tunnel its way through the horizon, ultimately
escaping to infinity as an observable quanta of  Hawking radiation.
Given this happenstance, the negative-energy partner must then tunnel inwards
until it terminates at the black hole singularity, thereby   
lowering the  mass of the black hole.

\par
There has been a great deal of success at using this tunneling framework
--- primarily,
as a means of reproducing the expectant value of the Hawking temperature
 ---  for a large assortment of black hole (and de Sitter) spacetimes
({\it e.g.}, \cite{PW,elias1,allan1}).  However,
as  Angheben, 
Nadalini, Vanzo and Zerbini (ANVZ) have recently observed \cite{Van},
the relevant works did typically rely upon 
either unorthodox or patchwork coordinate systems.
With this (perhaps worrisome) point as a central motivation,
ANVZ  went on to reformulate  the
calculation so that it bypasses any such coordinate specifics.
Their updated version relies, rather, upon the {\it proper} spatial distance;
that is, a measure of distance which is, appealingly, a coordinate
invariant.
 
\par
Although ANVZ did manage to reproduce the Hawking temperature
for a large class of black hole   spacetimes,
they neglected the self-gravitational effects of
the radiating particle; opting
to work, as a matter of choice, 
with the classical geometry throughout.
Meanwhile, many earlier works have stressed the utility
of the tunneling paradigm for this very purpose; that is, for
incorporating the effects of the back-reaction on the background 
spacetime.  Hence, a quite natural generalization
of the ANVZ analysis would be to see if their formalism can be viably extended
into this quantum-gravitational
 regime.  Just such an extension is, in fact, the objective
of the current letter.

\par
Let us start here by recalling the main result of ANVZ; namely, 
their formulation  of the classical action ($I$)  that describes the
trajectory of a tunneling particle.
(The reader can, of course, consult  the cited work \cite{Van}
for  the technical details and explanations
leading up to this outcome.)
Assuming a static  black hole metric of the generic form
\be
ds^2 \;=\; -A(r) dt^2 \;+\; B^{-1}(r) dr^2 \;+\; C(r)h_{ij} dx^i dx^j\;,
\label{1}
\ee
and a tunneling particle of energy $E$,~\footnote{Note that
this is the energy as measured by a stationary observer at infinity. 
The rest mass of
the particle also enters into the original
calculation, but this input is rendered  inconsequential because of the
horizon red-shift.} they obtain for the action \cite{Van}
\be 
I={2\pi i\over \sqrt {A^{\prime}(r_H)B^{\prime}(r_H)}}E+({\rm real}\;
{\rm contribution})\;,
\label{2}
\ee
where a prime denotes a differentiation with respect to $r$, and
$r_H$ locates the (classical)  radius of the horizon. (We
will be assuming, for the sake of  simplicity, a black hole spacetime 
containing a
single  horizon. The discussion can be appropriately
generalized to other circumstances, as elaborated on
in \cite{Van}.)
On this basis, ANVZ were able to deduce a semi-classical
tunneling probability  of
\be 
\Gamma=\exp\left[-2\Im I\right]=\exp
\left[-{4\pi E \over \sqrt {A^{\prime}(r_H)B^{\prime}(r_H)}}\right]\;.
\label{3}
\ee
This tunneling probability nicely coincides with the usual
Boltzmann factor $\exp[-\beta E]$, as follows readily from
the identification of the  inverse Hawking temperature 
\cite{Haw},~\footnote{And this
identification follows, in turn,  from Euclidean
path-integral considerations \cite{GHaw}.}
\be
\beta= {4\pi\over\sqrt {A^{\prime}(r_H)B^{\prime}(r_H)}}\;. 
\label{4}
\ee

\par
As mentioned above, the ANVZ calculation purposefully neglects
 the back-reaction of the particle on the background spacetime. 
So one might well ask
as to  how such an effect  can then be incorporated.
To address this query, let us first point out
that the sensible presumption of  energy conservation has a very 
important implication
\cite{PW}.
 Namely, a particle 
of instantaneous energy $\omega$ will effectively ``see'' a spacetime
metric of the form 
\be
ds^2 \;=\; -A[r(M-\omega)] dt^2 \;+\; B^{-1}[r(M-\omega)] dr^2 \;+\; 
C[r(M-\omega)]h_{ij} dx^i dx^j\;,
\label{5}
\ee
where $r$ is now expressed explicitly as a function of
the conserved (ADM) black hole mass, and
$M$ is the  value of this mass prior to the tunneling event
of interest. Hence, as a first approximation, one might (naively)
suggest that equation (\ref{2}) should be rewritten  with the simple  
replacement
$r_H(M)\rightarrow r_H(M-E)$.
However, because of the quantum uncertainty principle,
it is unnatural  to expect that the black hole mass can jump,
 from $M$ to $M-E$, 
in such a  discontinuous manner.
Rather, quantum blurring will require a ``gradual''
transition (relative to whatever time scale is characteristic of
the radiation process); so that it is much more accurate to
replace $r_H(M)$ with $r_H(M-\omega)$ and then suitably  integrate over
$\omega$.
That is (neglecting the irrelevant real part and
distinguishing the corrected action by a subscript $q$),
\bea 
I_{q} & \equiv & \int_{0}^{E}  \left. I \right|_{M\rightarrow M-\omega}
 d\omega \nonumber \\
& = & \int_{0}^{E} 
 {2\pi i\over \sqrt {A_H^{\prime}(M-\omega)B_H^{\prime}(M-\omega)}}
d\omega \;,
\label{6}
\eea
where  a subscript $H$ is a reminder that we are evaluating this quantity
at the   black hole horizon.~\footnote{In general, the
location of the black hole horizon can be obtained from the defining relation 
$A_H(M-\omega)=B_H(M-\omega)=0$.}

This is all well and good, but the reader might justifiably wonder 
if such a (relatively simple) modification could correctly account 
for the (complicated) effect of
the back-reaction. We will now proceed to demonstrate that our 
adaptation is,
indeed,  appropriately formulated for just this purpose.

 First of all, let us, for the sake of convenience,
rewrite equation (\ref{6}) in a more concise form  [{\it cf}, equation
(\ref{4})],
\be 
I_{q} ={i\over 2} \int_{0}^{E} \beta(M-\omega) d\omega \;.
\label{7}
\ee
Now, with the reasonable assumption that $M>>E$ ({\it i.e.},
the black hole is much more energetic than any emitted particle),
we can Taylor expand to  obtain
\be 
I_{q} ={i\over 2}\beta(M) \int_{0}^{E} \left[1-\omega
{\partial_M\beta(M)\over \beta(M)} +{\cal O} (\omega^{2})\right]
 d\omega \;.
\label{8}
\ee
Integrating, we then have
\be 
I_{q} ={i\over 2}\beta(M) E \left[1-
{E\over 2}{\partial_M\beta(M)\over \beta(M)} +{\cal O} (E^{2})\right]
 \;.
\label{9}
\ee

\par
Next, let us call upon the first law of black hole mechanics,
or 
$\beta(M)=\partial_M S(M)\;$, with $S$ representing the 
Bekenstein--Hawking entropy \cite{Bek,Haw}
of the black hole in question.
It follows that
\be 
I_{q} ={i\over 2}\left[E\partial_M S(M) -
{E^2\over 2}\partial_M^2 S(M) +{\cal O} (E^{3})\right]
 \;,
\label{10}
\ee
which (by Taylor ``contracting'') can  succinctly be rewritten as
\be 
I_{q} = -{i\over 2}\left [S(M-E)-S(M)\right]\;.
\ee
\label{11}
Notably, the square brackets contain simply the change in Bekenstein--Hawking 
entropy due to the emission of a particle of energy $E$
from the black hole.

\par
Returning to equation (\ref{3}), we are now able to deduce a {\it 
quantum-corrected} tunneling probability of
\be 
\Gamma_q=\exp\left[-2\Im I_q\right]=\exp
\left[S(M-E)-S(M)\right] \;,
\label{12}
\ee
as would be expected from the viewpoint of statistical mechanics.
That is to say, in any reasonable quantum tunneling process,
it is natural to anticipate a probability of $\Gamma\sim \
\rho_{final}/\rho_{initial}=\exp\left[S_{final}-S_{initial}\right]$
(with $\rho$ indicating  the density of states), which is precisely
what we have found here. This observation (as well as our agreement
with earlier treatments; {\it e.g.}, \cite{PW})  substantiates
our {\it ansatz} for incorporating the effects of the back-reaction.
And, although we have focused on a single-horizon  spacetime,
we expect the same basic approach to persist for the
other cases considered in \cite{Van}.
\par
Finally, let us point out  (as was recently emphasized in \cite{Parik})
that the outcome $\Gamma\sim \exp(\Delta S)$  is indicative
of a unitary theory underlying the
process of black hole evaporation. Which is to say, once the back-reaction
effects have been included, there is no longer any reason to expect
a loss of information from the black hole universe. (This can
also be argued for on the basis of
an energy-dependent {\it effective} temperature~\footnote{To see
this energy dependence, first
consider that $2\Im I_q$ can  be identified with $\beta_{eff}E$.
Then, by way of equation (\ref{9}),
this effective (inverse) Hawking temperature is equivalent to
the usual static value {\it plus} an infinite power-series expansion in $E$.}
and, consequently, a non-thermal spectrum.)
Given that there is no consensus viewpoint on resolving
the so-called {\it information loss paradox} \cite{Rou}, it would 
certainly be beneficial
if this tunneling perspective  could be better understood.
Work on this matter is currently in progress.

\newpage

\section*{Acknowledgments}
%\par

Research for AJMM is supported  by
the Marsden Fund (c/o the  New Zealand Royal Society)
and by the University Research  Fund (c/o Victoria University).
The work  of ECV is financially supported by the PYTHAGORAS II 
Project ``Symmetries in Quantum and Classical Gravity''
of the Hellenic Ministry of National Education and Religions.

%\vspace*{20pt}

\end{document}